\documentclass[prb,showpacs,preprint,amsmath]{revtex4}
\usepackage{latexsym}
\usepackage{graphicx}

\newcommand{\be}{\begin{equation}}

\newcommand{\ee}{\end{equation}}

\newcommand{\reff}[1]{(\ref{#1})}

\newcommand{\giro}[1]{\stackrel{\circ}{#1}}

\begin{document}

\title{Entanglement  in the states of the Two-Rotors Model}

\author{ Fabrizio Palumbo}

\affiliation{
INFN Laboratori Nazionali di Frascati, 00044 Frascati, Italy\\
}

\begin{abstract}

The eigenfunctions  of the Two-Rotors Model are superpositions of states corresponding to precessions of  the rotors around  two orthogonal axes. In the application of the model to a system of particles such a structure becomes a coherent entanglement  of many particles. In Nuclear Physics such an entanglement  has not been directly confirmed. I show that it is possible to come to a definite conclusion about its existence by measuring the em transition probabilities for the J=3 member of the scissors mode rotational band and for higher excited states with intrinsic energy twice that of the scissors mode. The present results are relevant to  single domain magnetic nanoparticles.  

\end{abstract}


\maketitle

\section{Introduction}

 The Two-Rotors  Model (TRM) describes the dynamics of two rigid bodies rotating with respect to each other under an attractive force around their centers  of mass fixed at one and the same point. It was devised as a  model for deformed atomic nuclei, in which case  the rigid bodies represent the proton and neutron systems [\onlinecite{LoIu}].  The low lying excited  states predicted by  this model  were first observed ~[\onlinecite{Bohle}] in the  rare earth nucleus $^{156}Gd $, and then  in all deformed atomic nuclei~[\onlinecite{LoIu1}], and  were called scissors modes, see Fig.1.

 By analogy similar collective excitations were predicted in several other systems~[\onlinecite{Guer}] and as  it is well known they  have been clearly observed in Bose-Einstein condensates~[\onlinecite{Mara}]. Moreover an application of the TRM to the evaluation of the magnetic susceptibility of single domain magnetic nanoparticles stuck in rigid matrices has given results  compatible with a vast body of experimental data with an agreement  in some cases surprisingly good~[\onlinecite{Hata}].
 
  Fig.1, however, while very suggestive, does not give a complete  representation  of the TRM states, because   the TRM Hamiltonian has a double well potential and then at the classical level two states corresponding to the two minima. The present paper is devoted to  the investigation of the consequences of this feature. 
In order to describe the problem it is necessary  to define the model.  I assume the two rotors to have axial symmetry~[\onlinecite{Palu}].The TRM Hamiltonian is then
 \be
H =\frac{1}{2{\mathcal I}_1}  {\vec L}_1^2 +  \frac{1}{2{\mathcal I}_2} {\vec L}_2^2 +V \label{TRMH}
\ee
where ${\vec L}_1, {\vec L}_2  $ are the angular momenta,  ${\mathcal I}_1, {\mathcal I}_2$  the moments of inertia of  the two rotors with respect to the axes perpendicular to the symmetry axis  and $V$ is  the  potential interaction between them.  I assume the potential to be a function of the angle  between the axes of the rotors. Denoting 
this angle by $2 \theta$ 
\be
V= V(| \cos ( 2 \theta)| )\,.
\ee

    \begin{figure} 
  \begin{center}
    \begin{tabular}{cc}
    \includegraphics[width=5cm]{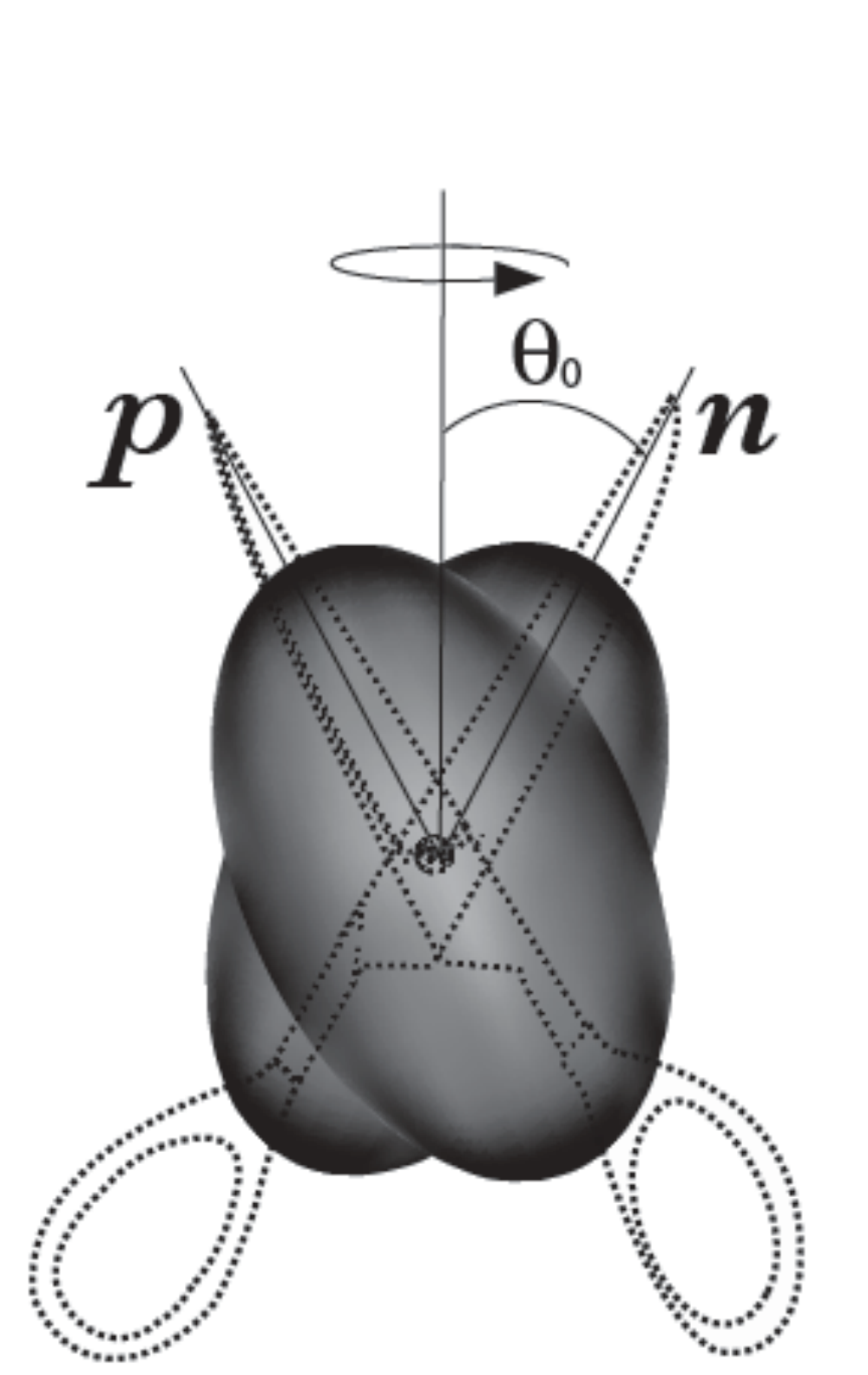}
    \end{tabular}
  \end{center}
 \caption{Scissors modes in the Two Rotors Model:  the proton (p) and neutron (n) rotors  precess around the bisector  of their axes. }
 
\end{figure}

 This potential is symmetric with respect to $ \theta= \pi/4$  and  has two degenerate minima at $\theta=0, \pi/2$. So at the classical level the axes of the rotors will vibrate with respect to one another around these values of $\theta$. At the quantum level the eigenfunctions will be superpositions of states describing these vibrations.  To be definite let us define  an intrinsic frame of axes $\xi, \eta, \zeta$ 
\be
{\hat \xi}=\frac{{\hat \zeta}_2 \times {\hat \zeta}_1}{ 2 \sin\theta},  \,\,\,\,\,
{\hat \eta}=\frac{{\hat \zeta}_2 - {\hat \zeta}_1}{2 \sin\theta}, \,\,\,\,\,
{\hat \zeta}=\frac{{\hat \zeta}_2 + {\hat \zeta}_1}{2 \cos \theta} \,.  \label{axes}
\ee
The eigenfunctions will be superpositions of states describing the precession of the proton and neutron axes   around  the $\zeta$- and  the $\eta$- axes~[\onlinecite{LoIu}]. Such superpositions  are constrained by the condition that  independent   inversions of the orientation of the proton and neutron  axes are not observable. In general such constraint should be imposed on the absolute value of the wave functions[\onlinecite{Palumbo}].
 In reference [\onlinecite{LoIu}], however,  they were enforced by requiring that the eigenfunctions, rather than their absolute value,  should be invariant under these inversions, and we will use here this restrictive requirement. As a consequence they result to  have the form schematically  represented in Fig. 2.

    \begin{figure} 
  \begin{center}
    \begin{tabular}{cc}
    \includegraphics[width=12cm]{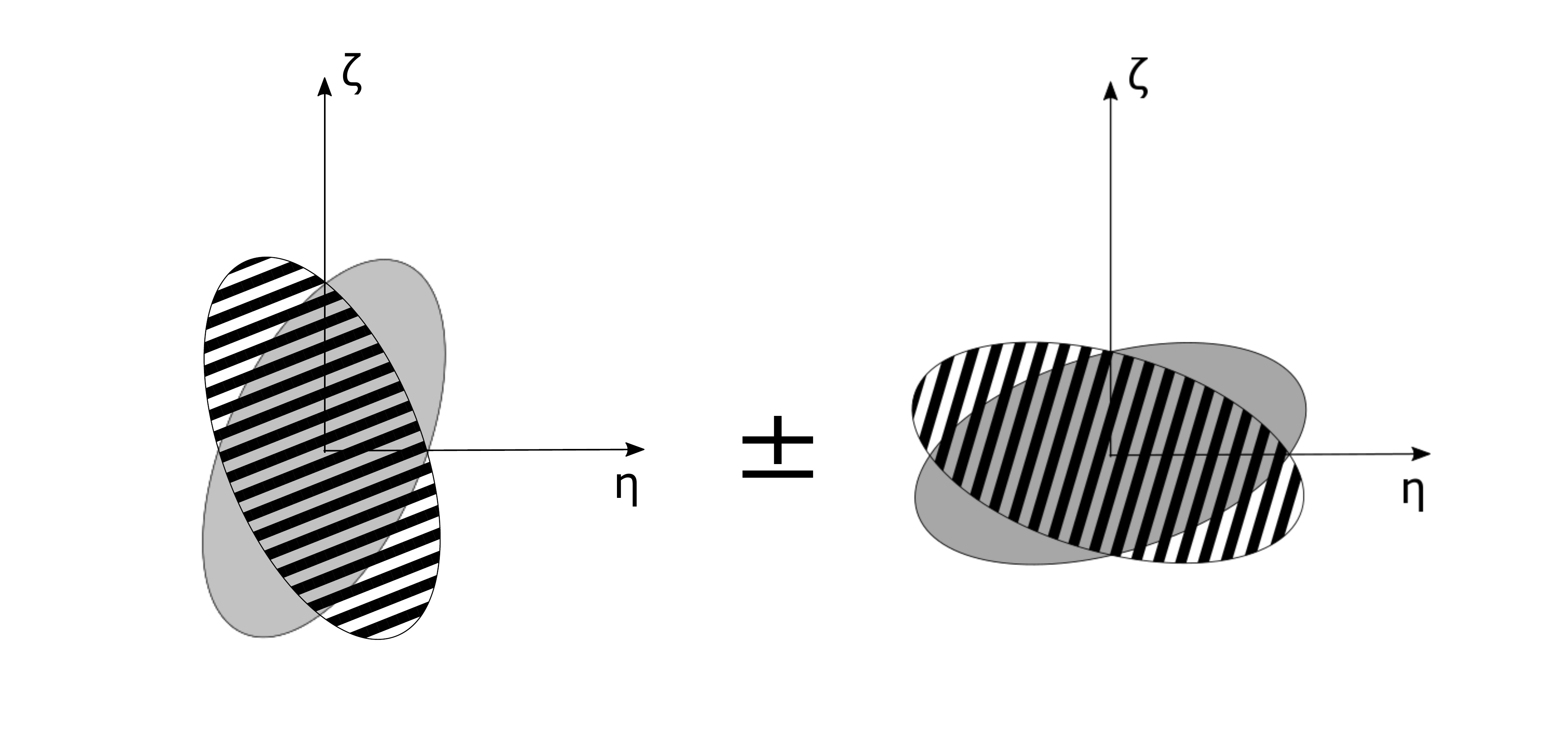}
    \end{tabular}
  \end{center}
 \caption{The TRM Hamiltonian has a double well potential, the two wells corresponding to the precession of the rotors axes around the $\zeta$- and $\eta$-axes of the intrinsic frame. The eigenfunctions therefore are {\it necessarily a  superposition of the states describing such precessions}. The   requirement that they be invariant  under inversion of the orientation of the neutron and proton axes, that is not observable, selects a definite  superposition for each value of the total angular momentum. }
\end{figure}

As far as I know the actual occurrence of such an entanglement  has never been directly confirmed  in Nuclear Physics.  I will discuss this feature of the TRM in detail and  I will show how entanglement can be observed in atomic nuclei by studying the  $J=3$ member  of the scissors mode rotational band and states with higher intrinsic energy. 

The determination of the eigenstates of the TRM requires the solution of the above mentioned constraint   that until now was worked out case by case. Here I present   a rather general and practical procedure to do it. In this way I find that the  solution for states with higher intrinsic energy  used in a previous investigation ~[\onlinecite{Palu1}]  is not unique, as I incorrectly assumed.

In Section II I  report the essentials of the TRM, in Section III I determine its eigenvalues and eigenstates,  in Section IV I discuss what we can learn about entanglement in atomic nuclei from the existing data and new possible experiments, in Section V I  compare with  other theoretical approaches, and finally in Section VIl  I present our conclusions, including a conjecture concerning  single domain magnetic nanoparticles [\onlinecite{Hata}]. In the Appendix I collect and derive some expressions of em operators. I set $\hbar=c=1$.

\section{ The Two-Rotors Model}

The TRM Hamiltonian acts on the direction cosines of the rotor axes ${\hat\zeta}_1, {\hat \zeta}_2$. 
These variables can be replaced  by the Euler angles  $ \alpha, \beta, \gamma$ that describe the orientation of the intrinsic frame  plus the angle $\theta$.
The correspondence $\{\zeta_1, \zeta_2\}= \{\alpha, \beta, \gamma, \theta\}$ is one-to-one and regular for $0< \theta<\pi/2$.  {\it  It is important to remember that this  whole range of $\theta$ is necessary for the  transformation to be one-to-one~[\onlinecite{DeFr}]}.

Because of the axial symmetry the wave functions must satisfy the constraints
\be
\zeta_1 \cdot L_1 \Psi = \zeta_2 \cdot L_2 \Psi =0
\ee
In order to get analytic results these (weak) constraints on the wave functions were replaced by (strong) constraints on the operators and solved in terms of
\begin{eqnarray}
L &=& L_1+L_2
\nonumber\\
S & =&L_1-L_2
\end{eqnarray}
where
\be
S_{\xi}= i \frac{\partial}{\partial \theta}\,, \,\,\,  S_{\eta}= - \cot \theta L_{\zeta} \,, \,\,\, S_{\zeta}= - \tan \theta L_{\eta}\,.
\ee
Using the above change of variables the TRM Hamiltonian  becomes  the sum of the rotational Hamiltonian of the two-rotors system as a whole plus an intrinsic Hamiltonian that in the reformulation of Ref.~[\onlinecite{DeFr}] reads
\be
H= \frac{{\vec  L}^2 }{2{\mathcal I}}+ H_{intr}
\ee
where
$
{\mathcal I}= {\mathcal I}_1 {\mathcal I}_2/
({\mathcal I}_1 + {\mathcal I}_2)
$
and
\begin{eqnarray}
H_{intr}  = \frac{1}{2{\mathcal I}}  \left[ \cot^2 \theta L_{\zeta}^2 + \tan \theta^2  L_{\eta}^2 - \frac{\partial^2}{\partial \theta^2}
-2 \cot(2\theta) \frac{\partial}{\partial \theta}\right] &&
\nonumber\\
+ \frac {{\mathcal I}_1 - {\mathcal I}_2}{4 {\mathcal I}_1 {\mathcal I}_2} 
\left[ -  \tan \theta  L_{\zeta}  L_{\eta} -  \cot \theta  L_{\eta}   L_{\zeta}      
+ i   L_{\zeta} \frac{\partial}{\partial \theta} \right] +V. && \label{Hintrinsic}
\end{eqnarray}
  This Hamiltonian is invariant under separate  inversions of the rotors  axes. To define the action of such operators I must write
 the unit length vectors ${\hat \zeta}_1,{\hat  \zeta}_2$ in terms of the intrinsic and global variables
 \be
{\hat \zeta}_1=-\sin \theta \, {\hat \eta} + \cos \theta \, {\hat \zeta}\,, \,\,\, 
{\hat \zeta}_2=\sin \theta \, {\hat \eta} + \cos \theta \, {\hat \zeta}\,.
\ee
Then the inversion operators  can be represented as 
\be
{\mathcal I}_{\zeta_1}= R_{\zeta}(\pi) R_{\xi}(\frac{\pi}{2}) R_{\theta}\,, \,\,\, \, {\mathcal I}_{\zeta_2}= 
R_{\eta}(\pi) R_{\xi}(\frac{\pi}{2}) R_{\theta}\,  \label {inversion}
\ee
where $ R_{\zeta}(\pi), R_{\eta}(\pi),  R_{\xi}(\frac{\pi}{2})$ are rotation operators around the intrinsic axes.

As I said it was assumed that such inversions should leave the wave functions  invariant. Invariance under separate inversions is equivalent to the conditions
\begin{eqnarray}
{\mathcal I}_{\zeta_1} {\mathcal I}_{\zeta_2}\Psi &= &\Psi \label{product}
\\
{\mathcal I}_{\zeta_1} \Psi &= &\Psi\,.  \label{single}
\end{eqnarray}

The range of $\theta$ can be separated into two regions
\be
s_I= s(\theta) s\left({\pi \over 4}-\theta\right), \,\, s_{II}= s\left({\pi \over 2}-\theta \right) \,s\left(\theta -  {\pi \over 4} \right), 
\ee
where $s(x)$ is the step function: $ s(x)=1, x>0$ and zero otherwise. The 2 regions are obtained from each other by  the reflection of $\theta$ with respect to ${\pi / 4}$. It is convenient to introduce the notation
\be
R_{\theta}f(\theta) =  f \left( {\pi \over 2} - \theta \right) = \giro{f}(\theta)   \label{reflection} \,,
\ee
so that $ \giro{s}_I = s_{II} $. With this notation  $\giro{V}=V$.

The second term of $H_{intr}$ is proportional to [\onlinecite{DeFr}] $ \theta_0 \, | {\mathcal I}_1 - {\mathcal I}_2| / (4 {\mathcal I}_1 {\mathcal I}_2 )$, where
 \be 
 \theta_0 = ({\mathcal I}C)^{-{ 1\over 4}}
 \ee
 is the zero point oscillation parameter. 
  It is therefore negligible for atomic nuclei (for which $|{\mathcal I}_1 - {\mathcal I}_2 | / (4 {\mathcal I}_1  {\mathcal I}_2  ) <<1 $ and  $\theta_0 \sim 0.1$)  but not for free nanoparticles  (for which $|{\mathcal I}_1 - {\mathcal I}_2 | / (4 {\mathcal I}_1  {\mathcal I}_2  ) \sim 1 $ and  $\theta_0 \le 1$). I think, however,  that the importance of the second term of $H_I$  for different moments of inertia is due to the fact that the intrinsic frame I choused is not a principal frame, namely a frame in which the tensor of the moment of inertia of the two-rotors system is diagonal. I conjecture that in a principal system the second term will  be small also for nanoparticles.  
  
  Neglecting the second term the TRM Hamiltonian becomes then  invariant also under the transformation
\be
R= R_{\xi} \left({\pi \over 2} \right) R_{\theta} \,. \label{R}
\ee
 Next I eliminate the linear derivative in the first term of $H_{intr}$  by  the transformation
\be
( U \Phi )(\theta)= { 1 \over  \sqrt{2 \sin(2 \theta)}} \,  \Phi'(\theta)\,. \label{transformation}
\ee
getting
 \begin{eqnarray}
H_{intr}' &=& U H_{intr} U^{-1}=  { 1 \over 2  {\mathcal I} } \Bigg[ - { d^2 \over d \theta^2 } -\left( 2+  \cot^2 ( 2\theta)\right)
\nonumber\\
&+& \cot^2 \theta \,  L_{\zeta}^2 + \tan^2 \theta   L_{\eta}^2  \Bigg] +V(\theta)\,. \label{Htransf}
\end{eqnarray}
At last  I assume that the wave functions have such a fast falloff (which is completely justified in the case of nuclei)   that I can perform  the harmonic approximation for the potential and the circular functions
\begin{eqnarray}
&&V  \approx {1\over 2} C \, \theta^2 \,,   \,\,\, S'_{\xi}= i  \,  \nabla_{\theta} \,, \,\,\,  S'_{\eta}= 0 \,, \,\,\, S'_{\zeta}= - \frac{1}{ \theta} L_{\eta}\,, \,\,\,\,
\mbox{in  region I}
\\
&&V  \approx {1\over 2} C \, {\giro{\theta}}^2 \,,   \,\,\, S'_{\xi}= -   i  \,  \nabla_{\giro{\theta}} \,, \,\,\,  S'_{\eta}= - \frac{1}{ \giro{\theta}} L_{\zeta} \,, \,\,\, S'_{\zeta}= 0 \label{S'}\,, \,\,\, \,
\mbox {in  region  II}
\end{eqnarray}
where
\be
\nabla_{\theta}= \frac{d}{d \theta} - \frac{1}{2 \theta}\,.
\ee
I then write accordingly
\be
H_{intr}' \approx H_I s_I + H_{II} s_{II}
\ee
where
\begin{eqnarray}
H_I&=&{1\over 2} \omega \left[ - {d^2 \over d x^2} + { 1\over x^2} \left( I_{\zeta}^2 - { 1\over 4} \right) + x^2 \right]\,, \,\,\, 
0\le x \le \frac{\pi}{4 \theta_0}
\nonumber\\
H_{II} &= &{1\over 2} \omega \left[ - {d^2 \over d \giro{x}^2} + { 1\over \giro{x}^2} \left( I_{\eta}^2 - { 1\over 4} \right) + \giro{x}^2 \right]\,, \,\,\,
0 \le \giro{x} \le \frac{\pi}{4 \theta_0}  
\end{eqnarray}
with
\be
 x= \frac{\theta} { \theta_0}\,, \,\,\,\,\, \giro{x}  \, = \frac{\giro{\theta}}{\theta_0}
\ee
\be
\omega = \sqrt{C \over {\mathcal I}}\,.
\ee
{\it The harmonic approximation makes  more evident   that \reff{Htransf} is  a double well Hamiltonian, implying that in stationary states \it the rotor axes oscillate simultaneously around  the $\zeta$- and $\eta$-axes}.

The eigenfunctions and eigenvalues of $H_I$ are~[\onlinecite{DeFr}]
\begin{eqnarray}
\varphi_{Kn}(x) &=& \sqrt{ { n! \over (n+K)! \, \theta_0}} \, x^{K+{1 \over2}} \, L_n^K\left(x^2 \right) e^{-{ 1\over 2} x^2}
\\
\epsilon_{nK}&=& \omega (2n +K +1)
\end{eqnarray}
where  $ L_n^K $ are Laguerre polynomials and the wave functions $\varphi_{Kn}$ are
normalized according to
\be
\int_0^{\infty} dx \, \left(\varphi_{Kn}(x)\right)^2 = { 1\over 2}\,.
\ee
Because in the harmonic approximation $\theta$ plays the role a a radius, I call $n$ the radial quantum number.

In general the  eigenstates  occur in doublets, whose  energy splitting can be estimated with  the WKB approximation
\be
\delta E \approx E \exp \int_{- \theta(E)}^{\theta(E)}(-|p(\theta)|) 
\ee
where $\theta(E)$  is the angle of inversion of the classical trajectory of energy $E$ and 
$p(\theta)$ its conjugate  momentum, $
|p|=\sqrt{ |2 {\mathcal I}(E-V) |} \approx |\sin \theta|/ \theta_0^2\,.
$
Because $\theta(E) \approx \theta_0$ for the states of interest
\be
\delta E \approx E \exp \left( - \frac{2}{\theta_0^2}  \right)\,.
\ee
For atomic nuclei in the rare earth region $\theta_0^2 \sim 0.01$ and such energy splitting is to all effects negligible, but the situation is different for nanoparticles.

\section{Eigenstates}

I write the eigenfunctions in the form
 \be
\Psi_{IMmn} = \sum_{K \ge 0}{\mathcal F}^I_{MK}(\alpha, \beta, \gamma) \Phi^I_{m Kn}(\theta) \label{standard}
\ee
where
\be
{\mathcal F}^I_{MK}= \sqrt{{2I+1}\over 16( 1 +\delta_{K0}) \pi^2 } \left( {\mathcal D}^I_{MK} +(-1)^I {\mathcal D}^J_{M-K}   \right). 
\ee
 $I,M,K$ are the nucleus angular momentum and its component on the $z$-axis of the laboratory frame and the $\zeta$-axis of the intrinsic frame,   and $m$ an  additional quantum number to be specified in the sequel. Because all the states I will consider have positive parity I will omit the parity quantum number. The combination of rotational matrices in the ${\mathcal F}$ is required by  the condition~\reff{product}. It remains to impose the condition~\reff{single}. 
 
 The eigenstates are normalized according to
\be
\int_0^{2\pi} d\alpha \int_0^{\pi}d \beta \sin \beta \int_0^{2 \pi }d\gamma \int_0^{{\pi \over 2}}d \theta \, |\Psi_{I M m n}|^2 =1\,.
\ee

The eigenstates of the Hamiltonian in region I are
\be
\Psi^{(I)}_{L,M,K,n} = {\mathcal F}^I_{MK}(\alpha, \beta, \gamma) \varphi_{Kn}(\theta)\,.
\ee
For each such eigenstate there is  in region II the degenerate eigenstate
\be
\Psi^{(II)}_{I,M,K,n}= {\mathcal G}^I_{M,K} \giro{\varphi}_{K,n}
\ee
where
\be
I_{\eta}^2 \, {\mathcal G}^I_{M,K}= K^2 {\mathcal G}^I_{M,K}\,.
\ee
The constraint~\reff{single} determines the their amplitudes in the total eigenfunction.

When I express the ${\mathcal G}^I_{M,K} $ in terms of the  $ {\mathcal F}^I_{MK} $ the total eigenfunctions  take the standard form \reff{standard}.
{\it Notice that in region I the eigenstates have a unique component of  $I_{\zeta}$, while  in region II  they have  all the  components of 
$I_{\zeta}$ appearing in ${\mathcal G}^I_{M,K} $. The quantum number $m$ is  the component of the total angular momentum on the $\zeta$-axis in region I. }
Even if each of the rotors  has axial symmetry, the two-rotors system   does not have it, so that the component of angular momentum along any intrinsic axis is not conserved, resulting in a superposition of  states with different $K$-quantum number. 

In order to impose the constraint \reff{single}  I must determine the action of  $R_{\eta}(\pi)$ and $R_{\xi}(\pi/2)$ on the $ {\mathcal F}^I_{M,K} $ and the ${\mathcal G}^I_{M,K} $. For any component of ${\hat I}_k, k=\xi, \eta, \zeta$ 
\be
\exp(i \alpha {\hat I}_k) = i \frac{{\hat I}_k}{I_k} \sin(I_k \alpha) + \cos(I_k \alpha)
\ee
so that
\begin{eqnarray}
\exp(i \pi {\hat I}_k) \psi_{I_k} &=& (-)^{I_k}\psi_{I_k}
\nonumber\\
\exp(i \pi/2 \, {\hat I}_k) \psi_{I_k} &=& \Big[ i \frac{{\hat I}_k}{I_k} \sin(I_k \pi/2) + \cos(I_k \pi/2)\Big] \psi_{I_k}
\end{eqnarray}
Notice that the  transformations in the last equation   are simpler  for $I_k$ even.

In order to find the action of $R_{\xi}(\pi/2)$ on the $ {\mathcal F}^I_{M,K} $ and the ${\mathcal G}^I_{M,K} $ I express these functions in terms of the eigenstates of $ {\hat I}_{\xi}^2$
\be
I_{\xi}^2 \, {\mathcal K}^I_{M,K}= K^2 {\mathcal K}^I_{M,K}\,.
\ee
Because, as noted above,  such an action is simpler for  even values of $K$ it is convenient to  express all the $ {\mathcal F}^I_{M,K} $ and the ${\mathcal G}^I_{M,K} $ for $K$ even and odd,  in terms of the $ {\mathcal K}^I_{M,K}$ with even $K$.

\section{The scissors mode rotational band}

For the discussion of entanglement it is necessary to separate the contributions coming from regions I and II. To this end I introduce the parameters $r_I, r_{II}$  that in the TRM take the values 
\be
 r_I= r_{II} =1\,, \,\,\,  \,\,\, \mbox{ in the TRM}\,.
\ee
A general feature is that  the intraband magnetic transition amplitudes vanish, because they are proportional to 
\be
<\varphi_{1,0}| \nabla_{\theta}| \varphi_{1,0}>=<\varphi_{1,0}|\frac{1}{\theta}| \varphi_{1,0}> = 0.
\ee

\subsection{The  band head}

The  band head, the scissors mode, is a  pure $K=1$ state. Its wave function and transition amplitude are well known~[\onlinecite{LoIu}] but are reported for the sake of completeness
 \be
\Psi_{1 M 1,0} = {\mathcal F}^1_{M1} \Phi^1_{1,1,0}
\ee
where
\be
\Phi^1_{1,1,0}=\varphi_{1,0} - \giro{\varphi}_{1,0}  \,.
\ee
The transition amplitude to the ground state is
\be
<\Psi_{1 M 1,0}|{\mathcal M}(M1;\mu)|\Psi_{0,0,0,0}> = \frac{i}{2 \sqrt 3} \, \frac{1}{\theta_0} {\mathcal M}(M1) C^{1M}_{001\mu} (r_I + r_{II})
\ee
where the expression of 
\be
{\mathcal M}(M1)= \sqrt \frac{3}{4 \pi} \, \frac{e}{2m}
\ee
is riderived in the Appendix.

\subsection{The J=2 member of the band}

The $J=2$ member of the band is also a pure $K=1 $ state, and its wave function and transition amplitude are also well known~[\onlinecite{LoIu}] but  are reported for the sake of completeness
\be
\Psi_{2 M 1,0} = {\mathcal F}^2_{M1} \Phi^2_{1,1,0}
\ee
where
\be
 \Phi^2_{1,1,0}= \varphi_{1,0} + \giro{\varphi}_{1,0}\,.
 \ee
Its transition amplitude to the ground state is
\be
 <\Psi_{2M,1,0}|{\mathcal M}(E2;\mu)|\Psi_{0,0,0,0}> = -i e \, Q_{20} \frac{1}{4} \theta_0 \,  \, C^{2M}_{002\mu} \, (r_I + r_{II})
\ee
where $Q_{20}$ is the quadrupole moment in the intrinsic frame.

\subsection{The J=3 member}

The wave function of the $J=3$ member is determined in the present paper. It can be written
 \be
\Psi_{3 M 1,0} =c \, {\mathcal F}^3_{M1} \, \varphi_{1,0} +  s \, {\mathcal G}^3_{M1} \giro{\varphi}_{1,0}\,, \,\,\, \,\,\,\,c^2+s^2=1
\ee
where
\be
{\mathcal G}^3_{M1}=\frac{1}{4}( {\mathcal F}^3_{M1} + \sqrt {15} \, {\mathcal F}^3_{M3} )\,.
\ee
The eigenfunctions of ${\hat I}_{\xi}^2$ with eigenvalues $0, 4$ respectively are
\begin{eqnarray}
{\mathcal K}^3_{M0}&=&\frac{1}{2 \sqrt 2}( {\sqrt 3 \, \mathcal F}^3_{M1} - \sqrt {5} \, {\mathcal F}^3_{M3}  )
\nonumber\\
{\mathcal K}^3_{M2}&=&\frac{1}{2\sqrt 2}(\sqrt 5  \, {\mathcal F}^3_{M1} + \sqrt {3} \, {\mathcal F}^3_{M3} )\,. 
\end{eqnarray}
Expressing  $ {\mathcal F}^3_{M1}  $ and $ {\mathcal G}^3_{M1} $ in terms of ${\mathcal K}^3_{M0} $ and ${\mathcal K}^3_{M2} $
and imposing  the  constraint~\reff{single} I get
\begin{eqnarray}
\Phi^3_{1, 1,0}&=& \varphi_{1,0} + \frac{1}{4} \giro{\varphi}_{1,0}
\nonumber\\
\Phi^3_{1,3,0}&=& \frac{\sqrt{15}}{4}\giro{\varphi}_{1,0}\,.
\end{eqnarray}
Written in the standard form~\reff{standard}
\be
\Psi_{3M1,0}= {\mathcal F}^3_1\Big(\varphi_{10} + \frac{1}{4}\giro{ \varphi}_{10} \Big) + \frac{\sqrt 15}{4} {\mathcal F}^3_3\giro{\varphi}_{10}\,.
\ee
One can  see how the intrinsic structure of the two-rotors system changes in the band with the angular momentum, with a strong departure from a rigid rotor.

The nonvanishing electromagnetic transition amplitudes are
\begin{eqnarray}
<\Psi_{3M1,0}|{\mathcal M}(E2;\mu)|\Psi_{2 M'1,0}> &=& e \, Q_{20} \frac{1}{\sqrt 7}  \, C^{3M}_{2M'2\mu}  <\varphi_{10}|\varphi_{10}> 
\Big( r_I - \frac{5}{4} r_{II}\Big)
\nonumber\\
<\Psi_{3M1,0}|{\mathcal M}(E2;\mu)|\Psi_{1 M'1,0}> &=&  e \, Q_{20} {\sqrt \frac{3}{7}}  \, C^{3M}_{1M'2\mu} < \varphi_{10}|\varphi_{10}>
\Big( 0.63 \, r_I  -  0.5 \, r_{II} \Big)
\nonumber\\
<\Psi_{3M1,0}|{\mathcal M}(M3;\mu)|\Psi_{0,0,0,0}> &=& i {\mathcal M}(M3) {\sqrt \frac{2}{7}} \, C^{3M}_{003\mu} \, C^{31}_{0031}  
< \varphi_{10}|\nabla| \varphi_{00}> \Big( r_I - \frac{1}{4}  \, r_{II} \Big)  \label{J=3}
\nonumber\\
\end{eqnarray}
where the expression of 
\be
{\mathcal M}(M3) = -   \frac{3}{20} {\sqrt \frac{42}{\pi}}  \,R_3^2  \Big[ 1- \frac{1}{3}\Big(\frac{R_1}{R_3}\Big)^3\Big] \Big[ 1- \Big(\frac{R_1}{R_3}\Big)^2\Big]
\ee
is derived in the Appendix.

\section{Overtones}

In previous papers~[\onlinecite{Palu1, Palu2}] I studied the states of intrinsic energy $2 E_{scissors}$,  called first overtones because of the harmonic approximation. I know that in general in Nuclear Physics we can trust collective models at most for the lowest excitation. Nevertheless I considered worth while investigating the first overtones  for two reasons.  First their excitation energy falls below the threshold for neutron emission and therefore  their   width is of purely electromagnetic nature, which  might make  their  existence plausible, in spite of the fragmentation of the scissors mode. Second, their electric quadrupole transition amplitude is of zero order[\onlinecite{Palu1}]  in $\theta_0$, and therefore much greater than  that of  the $J=2$ member of the scissors rotational band that  is of order $\theta_0$. 

I reconsider now these states by using the present method of solving the constraint~\reff{single}.

The state $I=0=m=0, n=1$ cannot be excited by electromagnetic radiation, and for this reason it was called  the elusive overtone~[\onlinecite{Palu2}].  The same is true for the state $I=1,m=0, n=1$. 

The states $I=2, m=0, n=1$ and $I=2, m=2, n=0$  are degenerate  and their wave functions  in regions  I and II are
\begin{eqnarray}
s_I \, \Psi_{2,M,0,1}&=&{\mathcal F}^2_{M0} \, \varphi_{0,1}
\nonumber\\
s_I \, \Psi_{2,M,2,0} &=&{\mathcal F}^2_{M2} \, \varphi_{2,0}\,.  
\end{eqnarray}

\begin{eqnarray}
s_{II} \, \Psi^{II}_{2 M 2,0} &=&{\mathcal G}^2_{M2} \, \giro{\varphi}_{2,0}
\nonumber\\
s_{II} \, \Psi^{II}_{2 M 0,1} &=&{\mathcal G}^2_{M0} \, \giro{\varphi}_{0,1}   
\end{eqnarray}
where
\begin{eqnarray}
{\mathcal G}^2_{M0}&=&\frac{1}{2}( {\mathcal F}^2_{M0} + \sqrt {3} \, {\mathcal F}^2_{M2}     )
\nonumber\\
{\mathcal G}^2_{M2}&=&\frac{1}{2}( {\mathcal F}^2_{M0} -  {\mathcal F}^2_{M2}     )\,.
\end{eqnarray}
The eigenfunctions of ${\hat I}_{\xi}^2$ with eigenvalues $0, 4$ respectively are
\begin{eqnarray}
{\mathcal K}^2_{M0}&=& \frac{1}{2 }( { \mathcal F}^2_{M0} - \sqrt {3} \, {\mathcal F}^2_{M2}     )
\nonumber\\
{\mathcal K}^2_{M2}&=&\frac{1}{2} (\sqrt 3  \, {\mathcal F}^2_{M0} +  {\mathcal F}^2_{M2}     )\,.
\end{eqnarray}
Expressing the $ {\mathcal F}^2_{MK}  $ and $ {\mathcal G}^2_{MK} $ in terms of ${\mathcal K}^2_{M0}$ and ${\mathcal K}^2_{M2} $
and imposing the constraint~\reff{single} I find 
\begin{eqnarray}
\Phi^2_{0, 0,1}&=& \varphi_{0,1} -  \frac{1}{2} \giro{\varphi}_{0,1}
\nonumber\\
\Phi^2_{ 0, 2,1}&=& -  \frac{\sqrt 3}{2} \giro{\varphi}_{0,1}
\nonumber\\
\Phi^2_{2, 0, 0}&=& - \frac{\sqrt{3}}{2}\giro{\varphi}_{2,0} 
\nonumber\\
\Phi^2_{2,  2, 0}&=& \varphi_{2,0} + \frac{1}{2} \giro{\varphi}_{2,0}\,.  \label{general}
\end{eqnarray}
The  state $\Psi_{2 M 0, 1} $ can be regarded as a member of the rotational band over the elusive overtone  $ \Psi_{0,0, 0,1}$.  Its quadrupole transition amplitude to the ground statae vanishes. Because, as the bandhead,  it cannot be excited from the ground state, I will not discuss it any further (even though if reached from above, it could decay to the scissors mode).

The  nonvanishing  electromagnetic transition amplitudes of the state  $\Psi_{2 M 2,0} $ are
\begin{eqnarray}
<\Psi_{2M2,0}|{\mathcal M}(E2;\mu)|\Psi_{0,0,0,0}> &=& e \, Q_{20} \frac{1}{4} \, \sqrt \frac{3}{10} \, C^{2M}_{002\mu} \, r_{II}
\nonumber\\
<\Psi_{2M2,0}|{\mathcal M}(M1;\mu)|\Psi_{1,M' 1,0}> &=&i {\sqrt \frac{3}{5}} \, \frac{1}{4 \, \theta_0} {\mathcal M}(M1) C^{2M}_{1M'1\mu}
( r_I  +  r_{II} )\,.
\end{eqnarray}
The transition strengths are
\begin{eqnarray}
&& B(E2)\uparrow_{\mbox{overtone}} = {1 \over   32 \, \theta_0^2} \, \frac{4  r_{II}^2}{ (r_I + r_{II})^2} \, B(E2)\uparrow_{\mbox{scissors}} \label{BE2}
 \nonumber\\
&&  B(M1; \mbox{overtone} \rightarrow \mbox{scissors} ) = {1 \over 7} \, B(M1)\uparrow_{\mbox{scissors}}\,. \label{BE2}
\end{eqnarray}
In  the quoted  investigation of overtones~[\onlinecite{Palu1}]  I did not find the present~\reff{general}, most general solution of the constraint~\reff{single}, but the particular  solution
\be
\frac{1}{{\sqrt 2}}  \, ( \Psi_{2 M 2,0}  +  \Psi_{2 M 0,1})
\ee
which was incorrectly assumed to be unique, and the em transition amplitudes were evaluated accordingly.  I notice that with the TRM values of the parameters $r_I, r_{II}$, the electric quadrupole transition strength of the overtone  $\Psi_{2M2,0} $ is  a factor $ 2$ larger than that of the above state while the magnetic dipole transition strength  is a factor $4/7$ smaller.

\section{ Entanglement}

The TRM gives distinctive predictions that should  enable us to reach a definite conclusion concerning the existence of entanglement  in atomic nuclei.

To be definite I compare the predictions of the  TRM with those of a Reference Model that does not have entanglement. This Reference Model is what is often regarded to be {\it the }  TRM  as derived from microscopic models~[\onlinecite{Diep, Balb, Rich}]. It is the  intrinsic Hamiltonian  $H_I$ with the understanding that it acts on intrinsic wave functions defined and normalized in the whole range $0< \theta < \pi/2$. 
The Reference Model  has the same eigenvalues as the TRM. Unlike the TRM it has axial symmetry and  obviously  describes a precession around the $\zeta$-axis only.  Its eigenfunctions  are pure $K$-states and can be obtained from the eigenfunctions of the TRM setting  $\giro{\varphi}_{Kn}=0$ and  $<\varphi_{Kn}|\varphi_{Kn}>=1 $.  The transition amplitudes can be obtained from the expressions relative to the TRM by setting
\be
r_I=2\,, \,\,\, r_{II}=0\,, \mbox{ in the Reference Model}\,.
\ee

I discuss the entanglement separately for the different states.

\subsection{Entanglement in the scissors  rotational band}

\subsubsection{Entanglement in the J=1,2 members of the band}

  The em transition amplitudes of these states are the same in the TRM and in the Reference Model, because they are proportional to $r_I + r_{II}$, a quantity that takes the same value in both models. Therefore   we cannot learn anything about   entanglement from their comparison with experiment.

\subsubsection{Entanglement in the J=3 member of the band}

Let us denote  by $RM,TRM$ the transition amplitudes for the Reference Model and the TRM respectively. Then
\begin{eqnarray}
<\Psi_{3M1,0}|{\mathcal M}(E2;\mu)|\Psi_{2 M'1,0}>^{R M}&=& - 8 <\Psi_{3M1,0}|{\mathcal M}(E2;\mu)|\Psi_{2 M'1,0}> ^{TRM}
\nonumber\\
<\Psi_{3M1,0}|{\mathcal M}(E2;\mu)|\Psi_{1 M'1,0}>^{RM} &=& 9.7 <\Psi_{3M1,0}|{\mathcal M}(E2;\mu)|\Psi_{1 M'1,0}>^{TRM}
\nonumber\\
<\Psi_{3M1,0}|{\mathcal M}(M3;\mu)|\Psi_{0,0,0,0}>^{RM} &=& \frac{8}{3} <\Psi_{3M1,0}|{\mathcal M}(M3;\mu)|\Psi_{0,0,0,0}>^{TRM}\,.
\end{eqnarray}
One can see that the amplitudes for decay of the $J=3, m=1,n=0$ state to the lower members of the band  are depressed by large   factors  in the TRM with respect to the Reference Model.   Eqs.\reff{J=3}  show that this is due to  destructive interference between the contributions from regions I and II. The difference in strengths is so large that if this member of the band can be observed one should be able to reach a definite conclusion about entanglement.

\subsection{Entanglement in the first overtones}

The electric quadrupole amplitude for decay of the  overtone $\Psi_{2M,2,0}$ to the ground state 
 {\it vanishes in the absence of entanglement}. The relation between the magnetic dipole transition amplitudes in the  Reference Model and the TRM is
\be
<\Psi_{2M2,0}|{\mathcal M}(M1;\mu)|\Psi_{0,0,0,0}>^{RM}= 1.7<\Psi_{2M2,0}|{\mathcal M}(M1;\mu)|\Psi_{0,0,0,0}>^{TRM}\,.
\ee
Observation of the magnetic transition in the absence of the electric decay would give strong support to the absence of entanglement. Obviously on the contrary, observation of both transitions with the strengths~\reff{BE2} would be evidence in favor of it. 

In a recent experiment  the deformed nucleus $^{156}$Gd, where the scissors mode has been discovered initially [\onlinecite{Bohle}], has been studied by a high resolution nuclear resonance fluorescence experiment at the S-DALINAC up to 7 MeV of excitation energy. " A single candidate with the following characteristics a) a ground state decay indicating a quadrupole radiation, and b) simultaneously a significant branch to the main fragment of the scissors mode at 3 MeV has not been found above the detection limit "[\onlinecite{Piet}].

For an  assessment of the realization in nature  of the first overtone and its entanglement it is crucial to put the above findings in relation with the present estimate of its   decay strength to the scissors mode. Indeed such a strength is not so large and in the comparison with experiment it should be reduced by a factor equal to the percentage of the total strength carried by the main fragment of the scissors mode.

\section{Other theoretical approaches} 

There is a copious literature on the scissors modes, in which however entanglement never appears explicitly. Therefore it is sufficient for me to examine schematically how could one investigate entanglement in the different approaches.  For this  purposes I can schematically divide them into two categories. 

In the first one,  following different procedures, one derives a collective Hamiltonian that has an eigenstate with the  quantum numbers of the scissors mode  and approximately the same excitation energy.  There can be however some important subtleties that I illustrate by two examples. 
One is provided by  the Interacting Boson Model~[\onlinecite{Iach}]. It has been shown~[\onlinecite{Diep}]   that in the coherent states approximation,  for small vibrations of the rotor axes around the $\zeta$-axis,  it  reproduces the intrinsic part  $H_I$ of the TRM Hamiltonian.
I think  that  the vibrations around the $\eta$-axis  are also present in the IBA Hamiltonian, and that in the coherent state approximation they should  provide the Hamiltonian $H_{II}$,  but  this  remains to be verified. I must notice, however, that in calculations done with the IBM one does not use the coherent state approximation, but rather other approximations assuming  explicit symmetries of the wave functions. In a comparison with the TRM one has to check whether and how the invariance under inversion  of the orientation of the rotors axes has been implicitly implemented, and whether  the assumed symmetries  imply, for instance,  axial symmetry, that would eliminate the entanglement altogether. 

Another relevant  example is the recent  analytical approach to rotational states~[\onlinecite{Rich}], in which the TRM Hamiltonian  has been derived in the form~\reff{TRMH}. This paper is especially interesting in our context, because in the derivation of the collective Hamiltonian, as far as I understand, entanglement has not been enforced explicitly, and then also the Hamiltonian of the Reference Model  should be a possible outcome. A clarification of this point is of the highest  consequence for a strict connection between a many-body Hamiltonian and the TRM. 

In conclusion one must be sure of which conditions concerning invariance under inversion  of the rotor axes
are explicitly or implicitly set  on the wave functions in the course of the derivation.

The second category includes model or microscopic calculations in which a collective state appears that can be interpreted as  the scissors mode.  The RPA for instance,  reproduces at a semiquantitative level the eigenvalues and the em strengths of the TRM  for scissors modes
~[\onlinecite{Suzu}].  A recent approach,  the Wigner Function Moments method~[\onlinecite{Balb}], also belongs  to this class.

In all the works belonging to this category, however,   to our knowledge the resulting collective modes have not been analyzed in relation to  the entanglement. 

All the theoretical approaches of which I am  aware are restricted to the lowest scissors excitation. This is justified by the fact that {\it in general
collective models in nuclear physics can be trusted at most for the first excited state. I notice, however, that this does not need to be an absolute rule, and indeed it  is not true   for all systems. For instance in the evaluation of the magnetic susceptibility of single domain magnetic nanoparticles using the TRM  all the excited states appear and contribute}~[\onlinecite{Hata}]. The important point  is whether the rotors actually behave as rigid bodies  at the  energy of the collective state of interest, namely whether the coupling between intrinsic and collective degrees of freedom is or is not important. A general criterion can be found in~[\onlinecite{Bohr}]. But for higher states  this point can  be more efficiently investigated in a constructive way, introducing in a many-body Hamiltonian a number of collective variables with an equal number of constraints
in order not to change the effective number of degrees of freedom.  In a variant of such a method one can avoid explicit  constraints that make the calculations akword by modifying the microscopic Hamiltonian in such a way as to push the spurious excitations associated with the redundant variables out the part of the spectrum one is interested in.   Such a method has been used long ago to enforce translational invariance~[\onlinecite{Palu3}] in shell model calculations and exploited to introduce collective rotations~[\onlinecite{Sche}]. The latter application might be extended to the physics of the TRM by introducing the collective  variable $\theta$ in addition to the Euler angles.

 \section{Conclusion}
 
The  wave functions of the TRM have a peculiar  entanglement. In applications of the model to many-body systems this becomes a coherent entanglement of many particles of which  I do not  know other examples.

In nuclear physics with the present  data  there is no   evidence in favor or against it, and  the only check I can envisage   is to compare  the mass density distribution of  the states in which the scissors mode is fragmented with that predicted by the  TRM.

I have shown, however, that  significant pieces of information   can be obtained from the study of  higher excited states.  I hope that a definitive assessment concerning the first overtone will come soon [\onlinecite{Piet}].  The other crucial investigation  concerns  the $J=3$ member of the scissors rotational band. If such a state is realized in nature and can be observed one has   enough distinctive predictions to identify it.

It is  interesting to consider the application of the TRM to single domain magnetic nanoparticles. These objects consist of a magnetic structure, called macrospin, that rotates with respect to a nonmagnetic lattice. They have been represented as a couple of rigid rotors, one associated with the nonmagnetic lattice, and the other one, with a spin attached, with the macrospin~[\onlinecite{Hata}]. The macrospin has usually two stable orientations antiparallel to each other, separated by an energy barrier. At finite temperature there is a finite probability for the magnetization to flip and reverse its orientation. The double well potential, at variance with the case of atomic nuclei in which it might appear an artifact, is in this case at the basis of the dynamics. There is a strong, even though  indirect evidence of the validity of the TRM for nanoparticles stuck in rigid matrices~[\onlinecite{Hata}].  I think that a direct check of the  entanglement predicted by  the TRM is possible by  measuring the magnetic susceptibility of free nanoparticles  at temperatures of the order of 1 K.

\appendix

\section{Electromagnetic  operators}

The magnetic multipole operator in the intrinsic frame is
\be
{\mathcal M}'(Ml, \mu)= \frac{e}{mc} \, \frac{1}{V_{nucleus}}  \int d{\vec r} \, S'_k \,  \frac{\partial}{\partial x_k} (r^l Y_{l \mu})
\ee
where $V_{nucleus} $ is the nuclear volume and the operators $S'_k$ in the intrinsic frame are given in   Eqs.~\reff{S'}. I found that the terms $S'_{\eta}, S'_{\zeta}$ do not contribute to the transitions of the states I consider, and I will ignore them.  Working out the above equation I then get the expression of the magnetic dipole (already well known) and octupole  operators in the laboratory frame
\begin{eqnarray}
{\mathcal M}(M1, \mu)&=& - {\mathcal M}(M1)  \frac{1}{{\sqrt 2}} (D^1_{\mu 1}-D^1_{\mu -1}) i 
( \nabla_{\theta} \, s_I -  \nabla_{\giro{\theta}} \,\,s_{II} )
\nonumber\\
{\mathcal M}(M3, \mu)&=& {\mathcal M}(M3) \frac{1}{{\sqrt 2}} (D^3_{\mu 1}-D^3_{\mu -1}) i ( \nabla_{\theta} \, s_I -  \nabla_{\giro{\theta}} \,\,s_{II} )
\end{eqnarray}
where
\begin{eqnarray}
{\mathcal M}(M1)&=& \sqrt \frac{3}{4 \pi} \, \frac{e}{2m}
\nonumber\\
{\mathcal M}(M3)&=& -   \frac{3}{20} {\sqrt \frac{42}{\pi}}  \,R_3^2  \Big[ 1- \frac{1}{3}\Big(\frac{R_1}{R_3}\Big)^3\Big] \Big[ 1- \Big(\frac{R_1}{R_3}\Big)^2\Big]
 \, \frac{e}{2mc}   \,.                                                                                                
\end{eqnarray}
 $R_3, R_1$  are the lengths of  the semiaxes of the ellipsoids. In the evaluation of transition amplitudes I will need the matrix elements
\be
<\varphi_{20}| \nabla_{\theta} | \varphi_{10}>= - \frac{1}{2\sqrt 2} \frac{1}{\theta_0}\,, \,\,\,\,\,
 <\varphi_{10}| \nabla_{\theta} |\varphi_{00}>= - \frac{1}{2} \frac{1}{\theta_0}\,.
\ee

The electric quadrupole operator in the laboratory frame was evaluated in~[\onlinecite{LoIu, Palu1}] 
\begin{eqnarray}
{\mathcal M}(E2,\mu) &=& e \, Q_{20}   \left[  
{\mathcal D}^2_{\mu 0}  \left(s_I - { 1\over 2} \, s_{II}  \right) +  {1\over 2} \sqrt {3 \over 2} \left( {\mathcal D}^2_{\mu 2} + {\mathcal D}^2_{\mu -2}   \right) s_{II} \right]
\nonumber\\
&& - i \frac{\sqrt 5}{2} e \, Q_{20}\, (\theta s_I + \giro{\theta} s_{II})\left( {\mathcal D}^2_{\mu 1} + {\mathcal D}^2_{\mu -1}   \right)
\end{eqnarray}
where  $e \, Q_{20}$ is the electric quadrupole moment in the intrinsic frame. Notice that the first line is of  zero order in $\theta$ while the second line is of order $\theta$. In the evaluation of transition amplitudes I will need the matrix element
\be
<\varphi_{20}|\varphi_{00}> = \frac{1}{2 \sqrt 2}\,.
\ee

\subsection *{Acknowledgment}

I thank   N. LoIudice for a discussion of the subject of the present paper and
N. Pietralla and A. Richter for a continuous correspondence and for keeping me informed about  their research related to  scissors modes.

\appendix

\end{document}